\documentclass[10pt,preprint2]{aastex}
\usepackage{wasysym}

\def\obliq{\varphi} 
\def\azimuth{\alpha} 
\def\inclination{i} 

\shorttitle{Transits of Fast-Rotating Stars}
\shortauthors{Jason W. Barnes}

\begin{document}

\title{Transit Lightcurves of Extrasolar Planets Orbiting Rapidly-Rotating Stars}
\author{Jason W. Barnes} \affil{Department of Physics} \affil{University of Idaho}
\affil{Moscow, ID 83844-0903} \affil{ResearcherID:  B-1284-2009} \email{jwbarnes@uidaho.edu}

\newpage

\begin{abstract} 

Main-sequence stars earlier than spectral type $\sim\mathrm{F6}$ or so are expected to
rotate rapidly due to their radiative exteriors.  This rapid rotation leads to an oblate
stellar figure.  It also induces the photosphere to be hotter (by up to several thousand
Kelvin) at the pole than at the equator as a result of a process called gravity darkening
that was first predicted by \citet{1924MNRAS..84..665V}.  Transits of extrasolar planets
across such a non-uniform, oblate disk yield unusual and distinctive lightcurves that can be
used to determine the relative alignment of the stellar rotation pole and the planet orbit
normal.  This spin-orbit alignment can be used to constrain models of planet formation and
evolution.  Orderly planet formation and migration within a disk that is coplanar with the
stellar equator will result in spin-orbit alignment.  More violent planet-planet scattering
events should yield spin-orbit misaligned planets.  Rossiter-McLaughlin measurements of
transits of lower-mass stars show that some planets are spin-orbit aligned, and some are
not.  Since Rossiter-McLaughlin measurements are difficult around rapid rotators, lightcurve
photometry may be the best way to determine the spin-orbit alignment of planets around
massive stars.  The \emph{Kepler} mission will monitor $\sim 10^4$ of these stars within its
sample.  The lightcurves of any detected planets will allow us to probe the planet formation
process around high-mass stars for the first time.

\end{abstract}

 
\section{INTRODUCTION}

Fifty-five transiting extrasolar planets have been discovered to date (see {\tt
http://exoplanet.eu/}).  Most of these planets orbit stars that have masses near
$1~\mathrm{M_\odot}$.  The primary reasons for this parent star mass bias for
transiting planets are twofold:  (1) stars of spectral type  later than K are too dim
to be caught in large numbers by wide-field transit surveys, and (2) stars
earlier than F have rotationally broadened spectral lines and inherent stellar noise
that make high-precision radial velocity follow-up impossible at present.

To address the early-star radial velocity problem, \citet{2007ApJ...665..785J} used
radial velocity to survey evolved high-mass stars that were formerly early-type
dwarfs when they were on the main sequence.  \citet{2007ApJ...665..785J} and
\citet{2008ApJ...675..784J} described 11 known planets around evolved stars with
$1.5~\mathrm{M_\odot} < M_* < 3.0~\mathrm{M_\odot}$, presumably former A stars.  Since
then 14 new planets around high-mass stars have been found:  NGC4349\#127b
\citep{2007A&A...472..657L}, 81 Cetus b \citep{2008PASJ...60.1317S}, NGC2423\#3b
\citep{2007A&A...472..657L}, 18 Delphinus b \citep{2008PASJ...60..539S}, HD17092b
\citep{2007ApJ...669.1354N}, 14 Andromedae b \citep{2008PASJ...60.1317S}, $\xi$
Aquilae b \citep{2008PASJ...60..539S}, HD81688b \citep{2008PASJ...60..539S}, HD173416b
\citep{2009LiuRIAA}, HD102272b and HD102272c \citep{2008arXiv0810.1710N}, 6 Lyncis b
\citep{2008PASJ...60.1317S}, HD5319b \citep{2007ApJ...670.1391R}, and OGLE2-TR-L9b
\citep{2008arXiv0812.0599S}.  The last of these, OGLE-2-TR-L9b, is the only radial
velocity planet whose host star has $M_* > 1.5~\mathrm{M_\odot}$ and lies on the
main sequence (spectral type F3V), and this planet is the only one that is known to
transit a massive star as well.

The \emph{Kepler} mission will discover many more transiting planets around early-type
stars, if they exist.  At least 10,000 main-sequence stars earlier than spectral type
F5 should be present in the \emph{Kepler} field, and will presumably be among the
mission's targets.  \emph{Kepler} will then be able to characterize the distribution of
planets with short periods around high-mass stars.  These close-in planets will
complement the far-out planets recently discovered around A dwarfs using direct imaging
\citep{ChristianMarois11282008,PaulKalas11282008}, and lead to a better understanding
of how planet formation varies with stellar mass.

Main sequence stars earlier than $\sim$mid-F spectral type, those with $M\gtrsim
1.5~\mathrm{M_\odot}$, are all expected to be fast rotators.  The structure of these
stars is such that they have radiative zones in their outermost layers, instead of a
convective zone near the surface like for our Sun.  The exterior convection in
later-type stars drives surface magnetic activity, which in turn drives strong stellar
winds that sap the star's angular momentum with time.  Early-type, exterior-radiative
stars retain their youthful high angular momenta, with some spinning at near the
breakup speed  \citep[\emph{e.g.}][and references therein]{2004sipp.book.....H}.  As a
result, early-type main sequence stars can be significantly oblate.  

The rotation induces an equator-to-pole gradient in the effective acceleration due to
gravity $g$ at the surface.  \citet{1924MNRAS..84..665V} showed that in such a case the
temperature of the star varies from equator to pole as well, a phenomenon called
gravity-darkening.  The Von Zeipel Theorem thus predicts that the flux emitted from the
surface of a rapidly rotating star is proportional to the local effective gravity. 
Thus the effect induces cooler temperatures (and hence lower emitted fluxes) at a
star's equator, and hotter temperatures at the poles.  The basic predictions of von
Zeipel theory were dramatically confirmed by recent optical interferometric
observations of Vega ($\mathrm{\alpha}$ Lyrae) 
\citep[][explaining residuals in earlier near-IR interferometry by \citet{2001ApJ...559.1147C}.]{2006Natur.440..896P} and Altair
($\mathrm{\alpha}$ Aquilae) \citep{2007Sci...317..342M}.  Gravity darkening is used
regularly to characterize close binary star systems from their lightcurves
\citep[\emph{i.e.}][]{2003A&A...402..667D}.  In binary systems that interact
gravitationally, tides can also reduce the effective gravity, resulting in gravity
darkening.

If planets around fast-rotating stars formed in-situ from the protostellar disk
or migrated to their present locations within that disk, then those planets
might be expected to orbit near their stars' equatorial planes.  If those
planets transit, then their orbital inclinations $\inclination$ are near
$90^\circ$ (using radial velocity teams' definition of $\inclination$ as the
angle between the planet's orbit pole and the plane of the sky).  The stellar
orbit pole would then be nearly coincident with the planet's orbit pole, giving
a stellar obliquity $\obliq$ of near $0^\circ$.  In transit, such a planet's
chord across its star's disk would be perpendicular to the projected stellar
rotation pole.  In this paper, I call these ``spin-orbit aligned" planets.  If
planets have experienced planet-planet scattering events in their past, however,
they might be expected to show significant spin-orbit misalignment
\citep{2008ApJ...686..603J}.

Rossiter-McLaughlin (R-M) measurements of the radial velocity of slowly-rotating stars
during planetary transits have been highly successful at determining spin-orbit
alignments.  Relatively low-precision R-M measurements of spin-orbit alignement were
made by \citet{2009arXiv0906.5605P} and \citet{2009A&A...498L...5M}. 
\citet{2006ApJ...653L..69W} and \citet{2007ApJ...665L.167W} made early measurements of
HD189733 and HD147506, showing them to be spin-orbit aligned. 
\citet{2008ApJ...686..649J} determined that HAT-P-1 is nearly aligned ($3.7^\circ \pm
2.1^\circ$).   \citet{2009arXiv0905.4727N} showed that HD17156 is nearly spin-orbit
aligned, with a misalignment of $10^\circ \pm 5.1^\circ$.  \citet{2007ApJ...667..549W}
determined a similarly misalignment of $12^\circ \pm 15^\circ$ in the HD149026 system,
consistent with spin-orbit alignment.  On the other hand, \citet{2009IAUS..253..508H}
show a striking spin-orbit misalignment of $70^\circ \pm 15^\circ$ in the XO-3 planetary
system.  Highly precise measurements from \citet{2009arXiv0907.2956T} show a tiny but
significant spin-orbit misalignment of $0.85^\circ\pm 0.3^\circ$ in the CoRoT-3 system. 
Unfortunately, Rossiter-McLaughlin measurements will likely be much more challenging
rapidly-rotating stars due to their high inherent radial velocity noise.

Because more massive stars rotate much faster than the Sun, the transit
lightcurves for the planets that \emph{Kepler} will discover around them will be
qualitatively and quantitatively different from those for planets orbiting
slowly-rotating stars.  \citet{1939ApJ....90..641R} considered this effect
for eclipsing binary stars, and here I consider the effects for transiting
planets.  Stellar oblateness will alter the times of transit ingress and egress,
and the overall transit duration, somewhat complementary to the effects of
oblate planets \citep{oblateness.2003,Seager.oblateness}.  When the star's spin
pole and the planet's orbit pole are aligned, the Von Zeipel effect will cause
systematic errors in radius determinations for the star and the planet, and will
lead to broadband color variations during transit.  If the stellar spin pole and
planetary orbit pole are not aligned, then bizarre transit lightcurves result
that can be used to constrain both the stellar spin pole direction and the
spin-orbit alignment.  In this paper I investigate the effect that a
fast-rotating star has on the lightcurves of transiting extrasolar planets in
preparation for the results expected from \emph{Kepler}.

\section{SYNTHETIC TRANSIT LIGHTCURVES} \label{section:lightcurves}

\subsection{Algorithm}

\begin{figure*}
\epsscale{2.1}
\plotone{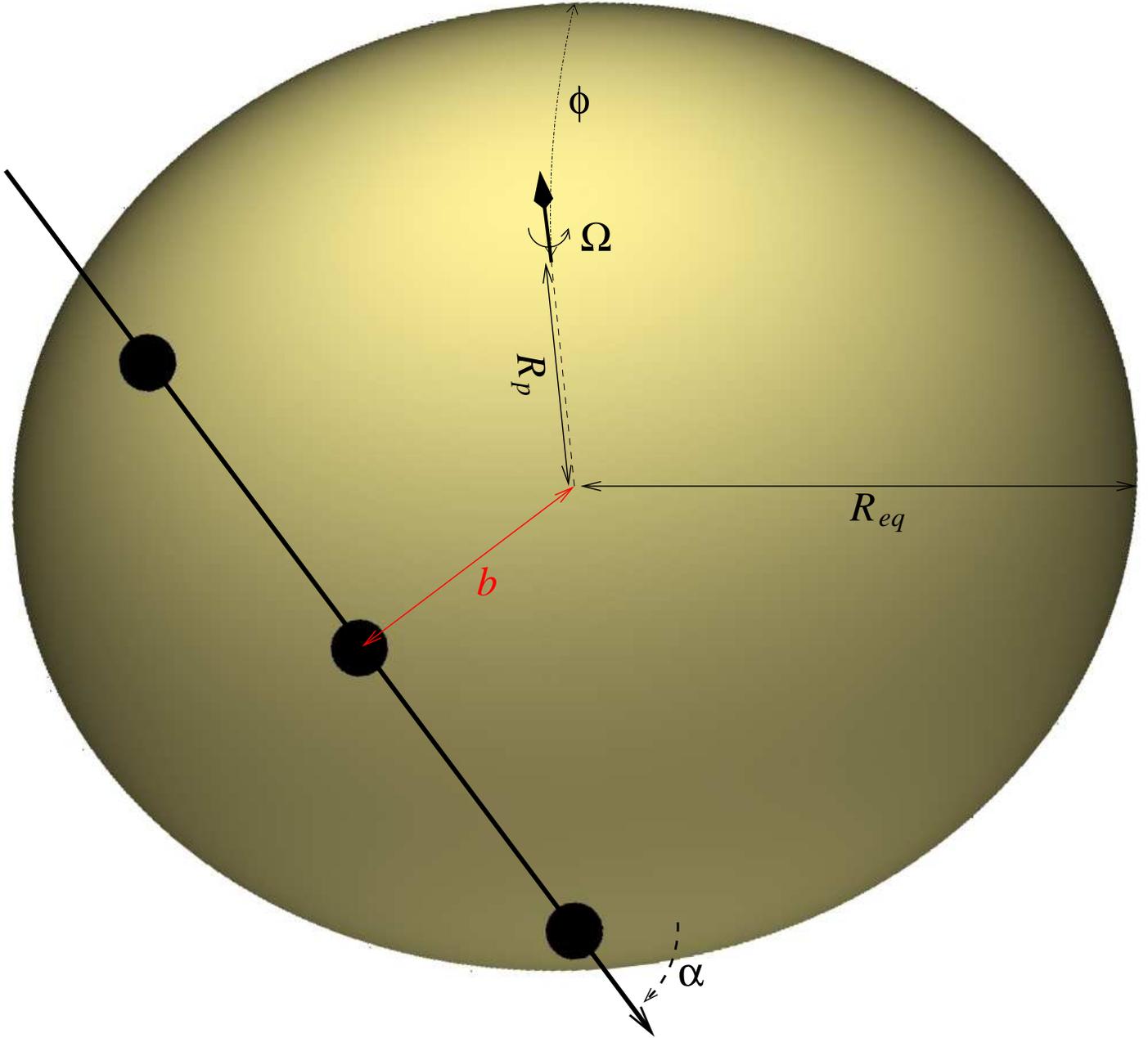}
\caption{ 
Schematic showing transit geometry along with some of the parameters referred to in the
text such as:  planet orbit azimuth $\azimuth$, transit impact parameter $b$, stellar
obliquity $\obliq$, stellar rotation rate $\Omega$, equatorial radius $R_{\mathrm{eq}}$,
and polar radius $R_\mathrm{p}$.
\label{figure:schematic}}
\end{figure*}

In order to generate synthetic transit lightcurves with fast-rotating stars, I modified
the algorithm originally developed for \citet{oblateness.2003} and extended in
\citet{2004ApJ...616.1193B} and \citet{2007PASP..119..986B}.  The algorithm numerically
integrates the total flux coming from the uneclipsed star, $F_0$, in polar coordinates
centered on the projected center of the star in the plane of the sky such that:
\begin{equation}
\label{eq:Finf}
F_0 = \int_{0}^{R_\mathrm{eq}} \int_{0}^{2\pi} I(r, \theta)~d\theta~dr
\end{equation}
where $R_\mathrm{eq}$ is the radius of the star at its equator (see Figure
\ref{figure:schematic} for a
schematic of some of the geometric variables), $r$ and $\theta$ are
measured from the stellar center and counterclockwise from the x-axis respectively,
and $I(r, \theta)$ is the star's intensity at point ($r$,$\theta$).
It then evaluates the apparent stellar flux
at time $t$, $F(t)$, relative to the out-of-transit flux $F_0$, by
subtracting the amount of stellar flux blocked by the planet from $F_0$:
\begin{equation}
\label{eq:Fblocked}
F_{blocked}(t) = \int_0^{R_\mathrm{eq}} \int_{0}^{2\pi} \Gamma(r,\theta,t) I(r,\theta)~d\theta~dr   ~\mathrm{,}
\end{equation}
and
\begin{equation}
\label{eq:integralalgorithm}
F(t) = \frac{F_0 - F_{blocked}}{F_0}~\mathrm{,}
\end{equation}
where $\Gamma(r,\theta,t)=1$ if the planet is blocking starlight at position $r$,
$\theta$ and time $t$, and $\Gamma(r,\theta,t)=0$ if not.

The difference in the case of fast-rotating stars is that the rotation induces these
stars' equators to bulge outward.  Hence the polar integral in Equations
\ref{eq:Finf} and \ref{eq:Fblocked} no longer properly accounts for the symmetry of
the problem.  I quantify this effect here as the star's oblateness, $f$, defined to
be $f\equiv\frac{R_\mathrm{eq}-R_\mathrm{pole}}{R_\mathrm{eq}}$ where $R_\mathrm{pole}$
is the star's radius along the rotational pole.  I assume throughout that the 
star's resulting
shape can be considered a MacLaurin spheroid.  The value
relevant for the integrals in Equations \ref{eq:Finf} and \ref{eq:Fblocked}, though,
is the effective oblateness $f_\mathrm{eff}$, which I define to be the apparent
oblateness of the star when projected into the plane of the sky.  The effective
oblateness is related to the actual oblateness by the stellar obliquity $\obliq$,
where $\obliq=0$ if the stellar rotation axis resides in the plane of the sky:
\begin{equation}\label{eq:feff}
f_\mathrm{eff} = 1 - \sqrt{(1-f)^2\cos^2\obliq + \sin^2\obliq}~.
\end{equation}
The relationship is not a simple cosine owing to three-dimensional geometry.  I will
show how this expression can be derived a few paragraphs below.

With $f_\mathrm{eff}$ in hand, it becomes straightforward to incorporate the stellar
asphericity.  In order to avoid complex and computationally-intensive elliptical
integrals, I execute a substitution for $r$ and $\theta$ in Equations \ref{eq:Finf}
and \ref{eq:Fblocked}.  I instead choose to integrate over $r'$ and $\theta'$, where
$r'$ and $\theta'$ are chosen so as to `pop' the star into a spherical shape in
$r'$-$\theta'$ space.  To do this, I first convert the true projected $r$ and
$\theta$ measured from the star's center into $x=r\cos\theta$ and $y=r\sin\theta$.  I
assume that the projected stellar rotation axis is parallel to the $y$-axis for
simplicity -- the true orientation will not be known, in general, but does not matter
since it does not affect the measured stellar flux.  I then let $x'=x$ and
\begin{equation}\label{eq:substitution}
y'=\frac{y}{(1-f_\mathrm{eff})}~,
\end{equation}
and then set $r'=\sqrt{x'^2+y'^2}$ and $\theta'=\mathrm{atan2}(y',x')$ where atan2 is the
C computer language arctangent function that returns a true 4-quadrant-capable angle 
from x and y values.  The substitution in Equation \ref{eq:substitution} works
equally well if you were to choose to integrate the stellar flux in cartesian xy
coordinates instead of the polar integral that I use.

Equations \ref{eq:Finf} and \ref{eq:Fblocked} now become:
\begin{equation}
\label{eq:Finfprime}
F_0 = (1-f_\mathrm{eff}) \int_{0}^{R_\mathrm{eq}} \int_{0}^{2\pi} I(r', \theta')~d\theta'~dr' 
\end{equation}
and
\begin{equation}
\label{eq:Fblockedprime}
F_{blocked}(t) = (1-f_\mathrm{eff}) \int_0^{R_\mathrm{eq}} \int_{0}^{2\pi}
\Gamma(r',\theta',t) I(r',\theta')~d\theta'~dr'  ~\mathrm{.}
\end{equation}
The $(1-f_\mathrm{eff})$ factor is introduced by the coordinate transformation: 
`popping' the oblate star out into a circle overestimates its projected area, and
hence the emitted flux, by $(1-f_\mathrm{eff})^{-1}$.  In the end the factor is
irrelevant.  When the results are plugged into Equation \ref{eq:integralalgorithm},
it drops out.  Hence the lightcurve generation algorithm as implemented does not 
use the factor $(1-f_\mathrm{eff})$ explicitly at all.

All that's left then is to determine $I(r',\theta')$ and then to integrate it. 
Obtaining $I(r,\theta)$ is straightforward but nontrivial.  I first break out the
stellar limb darkening from the normal emission, and assume blackbody radiation:
\begin{equation}\label{eq:Iwithlimbdarkening}
I(r',\theta')~=~B_\lambda(T(r',\theta')) L(r', \theta')
\end{equation}
where $B_\lambda$(T) is the blackbody function (or your desired stellar emission as a
function of temperature at a given wavelength), $T(r',\theta')$ is the temperature at
a given point on the stellar disk, and $L(r', \theta')$ is the stellar limb darkening
at that point.  It would be possible to incorporate a more realistic stellar emission spectrum rather
than to assume it to be a blackbody, but since rapidly-rotating stars are mostly of
early spectral types, the differences are not significant at the level of the present
investigation.

A modified version of the Von Zeipel theorem \citep{1924MNRAS..84..665V} determines
the stellar temperature $T$ at every point.  The temperature on the surface of a
rapidly rotating star is given by \citep{2009pfer.book.....M}
\begin{equation}\label{eq:vonZeipel}
T = T_\mathrm{pole} \frac{g^\beta}{g_\mathrm{pole}^\beta}
\end{equation}
where $g$ is the magnitude of the local effective surface gravity and $T_\mathrm{pole}$ and 
$g_\mathrm{pole}$
are the pole's temperature and surface gravity, respectively.  The value $\beta$ is
known as the gravity-darkening parameter.  Its nominal value is 0.25 for a purely
radiative star, as derived by von Zeipel; for our numerical calculations I use an
empirically measured $\beta$, as detailed below.  The local surface gravity vector
$\vec{g}$ has two terms:  one Newtonian, and one centrifugal:
\begin{equation}\label{eq:gravity}
\vec{g} = -\frac{GM_*}{R^2}\hat{r} + \Omega^2 R_\perp \hat{r_\perp}
\end{equation}
where $G$ is the universal gravitational constant, $M_*$ is the stellar mass,
$\Omega$ is the stellar rotation rate in radians per second, $R$ is
the distance from the star center to the point in question, and $R_\perp$ is the
distance from the star's rotation axis to the point in question.  The two hatted
symbols, $\hat{r}$ and $\hat{r_\perp}$ are unit vectors pointing to the point in
question from the stellar center and stellar rotation axis respectively.

So, then, to know $I$ you need to know $g$, and to know $g$ you need to know the
three-dimensional vector position of each point that you see on the star.  In this case
you already know $r'$ and $\theta'$, from which you can trivially derive $x$ and $y$,
the location of 2-d projection in the plane of the sky of the point of interest with
respect to the center of the star.  What is left then is to determine $z$.  This is
nontrivial.  

The geometrical constraint on $z$ is that its value must conform to the surface of an
oblate spheroid given $x$, $y$, the stellar radius $R_\mathrm{eq}$, and the
oblateness $f$.  This is easy enough when the stellar obliquity is zero, and thus
when the y-axis is parallel to the stellar rotation axis.  So I define a new set of
coordinate axes with the same origin as the x-y-z system, at the center of the star. 
However this new system is rotated in the y-z plane (\emph{i.e.} around the x-axis)
by an angle $\obliq$, the star's obliquity to the plane of the sky.  Call this the
$x_0$, $y_0$, $z_0$ system, where
\begin{eqnarray}
x_0 & \equiv & x~~,\nonumber\\ 
y_0 & \equiv & ~y\cos(\obliq) + z\sin(\obliq)~~,~~\mathrm{and}\nonumber\\
z_0 & \equiv & -y\sin(\obliq) + z\cos(\obliq)~~.\label{eq:xyz0}
\end{eqnarray}
In this new obliquity-rotated coordinate system, the surface of the star's 
photosphere follows
\begin{equation}\label{eq:ellipsoid}
x_0^2+\frac{y_0^2}{(1-f)^2}+z_0^2=R_\mathrm{eq}^2~.
\end{equation}
Plugging the definitions of $x_0$, $y_0$, and $z_0$ from Equations \ref{eq:xyz0}
into Equation \ref{eq:ellipsoid}, I solve for $z$ in
terms of a known $x$ and $y$.  The solution to the resulting quadratic is
\begin{equation}\label{eq:zfromxy}
z = \frac{-2y(1-(1-f)^2)\sin{\obliq}\cos{\obliq}+\sqrt{d}}
{2((1-f)^2\cos^2\obliq+\sin^2\obliq)} 
\end{equation}
where the determinant $d$ is
\begin{eqnarray}\label{eq:determinant}
\lefteqn{d \equiv} &&
4y^2(1-(1-f^2))^2\sin^2\obliq\cos^2\obliq + \\
&&-4((\cos^2\obliq(1-f)^2+\sin^2\obliq)\times \nonumber \\
&&((y^2\sin^2\obliq-R_\mathrm{eq}^2+x^2)(1-f^2)+y^2\cos^2\obliq)) \nonumber~.
\end{eqnarray}
I choose the positive root of the determinant as the negative root represents the
invisible second interception of the line-of-sight with the photosphere that occurs
on the back side of the star as seen from Earth.  Equation \ref{eq:feff} above can be
derived by  setting the determinant $d$ equal to zero to establish the outer edge of
the star's disk as seen from Earth, and then solving for the proper $f_\mathrm{eff}$
to reproduce that disk.

I now have all of the necessary parameters to compute the flux coming from each point
on the star.  To do so, at each $x$, $y$ point use Equation \ref{eq:zfromxy} to get
$z$, and then plug $z$ into Equations \ref{eq:xyz0} to get $x_0$, $y_0$, and $z_0$.  Do
the vector addition to get $\vec{g}$ from Equation \ref{eq:gravity} in $x_0$, $y_0$,
$z_0$ space where
\begin{eqnarray*}
\vec{R} &\equiv &\frac{x_0}{R}\hat{i_0}+\frac{y_0}{R}\hat{j_0}+\frac{z_0}{R}\hat{k_0} ~~~~,\\
\vec{R_\perp} &\equiv& \frac{x_0}{R_\perp}\hat{i_0}+\frac{z_0}{R_\perp}\hat{k_0}~~~~,\\
R & \equiv & \sqrt{x_0^2+y_0^2+z_0^2} \\
R_\perp & \equiv & \sqrt{x_0^2+z_0^2} 
\end{eqnarray*}
Then plug $g\equiv|\vec{g}|$ into Von Zeipel's Equation (Equation \ref{eq:vonZeipel})
to get $T$, and then derive a flux from $T$ using a blackbody curve or your choice of
a more sophisticated emitted flux.

\subsection{Parameters}

In order to generate appropriate and representative lightcurves that can be used for
comparison to transits yet undiscovered, I calculate all transit lightcurves as if the
parent star were Altair ($\mathrm{\alpha}$ Aquilae).  The true host stars for
\emph{Kepler}-detected transiting planets will show varying stellar masses, polar
temperatures, radii, and rotation rates.  I elect to use Altair because I think that
its spectral type (A7V) is broadly representative of the majority of the expected
fast-rotating stars in the \emph{Kepler} sample, and because its parameters are
well-characterized by interferometric imaging \citep{2007Sci...317..342M}.  Not all
sets of parameters produce physically plausible stars; I avoid non-physical combinations
by only using this one set of known stellar values.

The specific stellar parameters that I use are  $M_*=1.8~\mathrm{M_\odot}$
\citep{2006ApJ...636.1087P},  $R_\mathrm{eq}=2.029~\mathrm{R_\odot}$, 
$T_\mathrm{pole}=8450~\mathrm{K}$, $\beta = 0.190$, $f=0.1947$, and a stellar rotation
period of 8.64 hours 
\citep[all as directly measured for Altair by][]{2007Sci...317..342M}.  For the planet, I assume $R_p=\mathrm{R_{Jup}}$ and an
orbit semimajor axis of $0.05~\mathrm{AU}$ (corresponding to a period of 3.04 days) for
familiarity with the lightcurves of known transiting hot Jupiters.  Transit lightcurve
shapes are invariant with orbit period; only the timescale changes.  Hence the curves
that I show here can be converted for different-period planets by stretching the x-axis.

In fitting I adjust for the $c_1$ and $c_2$ limb darkening coefficients outlined in
\citet{2001ApJ...552..699B}.  I generate the synthetic lightcurves using $c_1=0.640$ and
$c_2=0.0$.  I also assume a monochromatic observation at 0.51 microns wavelength except where
otherwise noted.

\section{SPIN-ORBIT ALIGNED}

\begin{figure*}
\epsscale{2.1}
\plotone{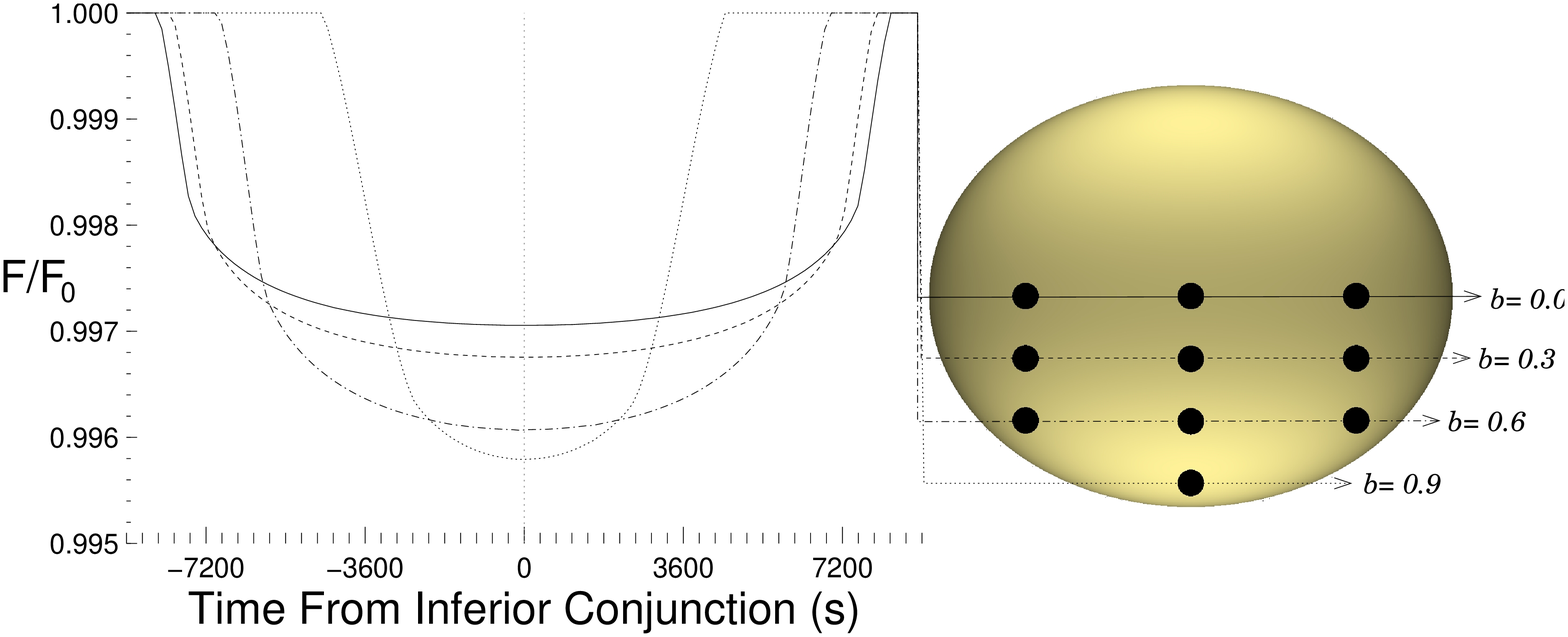}
\caption{Synthetic lightcurves for transiting 1 $R_{Jup}$ in a spin-orbit aligned $0.05~\mathrm{AU}$
orbit around an Altair-like star are plotted.  The four curves correspond to planets with
transit impact parameters of $b=0.0~R_{pole}$ (solid), $b=0.3~R_{pole}$ (dashed),
$b=0.6~R_{pole}$ (dot-dashed), and $b=0.9~R_{pole}$ (dotted).  The curves' shapes are
indistinguishable from transits of slow-rotating, spherical stars using different parameters.
Transits nearer to the pole are deeper because the stellar photosphere is hotter there due to
the Van Zeipel effect.
\label{figure:q0.a0}}
\end{figure*}

\begin{deluxetable}{|ccr|ccccl|}
\tablecaption{Best-fit transit parameters assuming spherical star for various transit
geometries.\label{table:fits}}
\tablewidth{0pt}
\tablehead{
\colhead{Stellar} &
\colhead{Planet's} &
\colhead{Transit} &
\colhead{Best-} &
\colhead{fit} &
\colhead{ Parameters}  \\
\colhead{Obliquity} &
\colhead{Azimuth} &
\colhead{Impact} &
\colhead{} &
\colhead{} &
\colhead{} &
\colhead{Limb}  &
\colhead{Limb} \\
\colhead{($\obliq$, $^\circ$)} &
\colhead{($\azimuth$, $^\circ$)} &
\colhead{Parameter} &
\colhead{$R_*$} &
\colhead{$R_p$} &
\colhead{Inclination} &
\colhead{Dark 1}  &
\colhead{Dark 2}  \\
\colhead{}&
\colhead{} &
\colhead{($b$, $\mathrm{R_{pole}}$)} &
\colhead{$\mathrm{R_\odot}$} &
\colhead{$\mathrm{R_{Jup}}$}&
\colhead{($\inclination$, $^\circ$)} &
\colhead{$c_1$}  &
\colhead{$c_2$}  
}\startdata
\hline
$0^\circ$   &   $0^\circ$  &  $0.0~\mathrm{R_{pole}}$  & $2.050~\mathrm{R_\odot}$  &
		$0.999~\mathrm{R_{Jup}}$   &   $88.23^\circ$   &  0.612 &\\
$0^\circ$   &   $0^\circ$  &  $0.3~\mathrm{R_{pole}}$  & $2.014~\mathrm{R_\odot}$  &
		$1.034~\mathrm{R_{Jup}}$   &   $86.95^\circ$   &  0.621 &\\
$0^\circ$   &   $0^\circ$  &  $0.6~\mathrm{R_{pole}}$  & $1.970~\mathrm{R_\odot}$  &
		$1.141~\mathrm{R_{Jup}}$   &   $84.07^\circ$   &  0.626 &\\
$0^\circ$   &   $0^\circ$  &  $0.9~\mathrm{R_{pole}}$  & $1.937~\mathrm{R_\odot}$  &
		$1.291~\mathrm{R_{Jup}} $   &   $80.94^\circ$   &  0.582 &\\ \hline
$30^\circ$  &   $0^\circ$  & $-0.9~\mathrm{R_{pole}}$  & $1.91~\mathrm{R_\odot}$  &
		$1.27~\mathrm{R_{Jup}}$   &   $81.70^\circ$   &  0.826  & 1.000 * \\
$30^\circ$  &   $0^\circ$  & $-0.6~\mathrm{R_{pole}}$  & $2.02~\mathrm{R_\odot}$  &
		$1.24~\mathrm{R_{Jup}}$   &   $84.01^\circ$   &  0.834  & 1.000 * \\
$30^\circ$  &   $0^\circ$  & $-0.3~\mathrm{R_{pole}}$  & $2.08~\mathrm{R_\odot}$  &
		$1.16~\mathrm{R_{Jup}}$   &   $85.93^\circ$   &  0.805  & 1.000 *\\
$30^\circ$  &   $0^\circ$  &  $0.0~\mathrm{R_{pole}}$  & $2.04~\mathrm{R_\odot}$  &
		$1.02~\mathrm{R_{Jup}}$   &   $88.70^\circ$   &  0.723  & 1.000 \\
$30^\circ$  &   $0^\circ$  &  $0.3~\mathrm{R_{pole}}$  & $2.02~\mathrm{R_\odot}$  &
		$1.01~\mathrm{R_{Jup}}$   &   $89.97^\circ$   &  0.726  & 0.859 \\
$30^\circ$  &   $0^\circ$  &  $0.6~\mathrm{R_{pole}}$  & $2.123~\mathrm{R_\odot}$  &
		$0.987~\mathrm{R_{Jup}}$   &   $83.05^\circ$   &  0.465 &\\ 
$30^\circ$  &   $0^\circ$  &  $0.9~\mathrm{R_{pole}}$  & $2.143~\mathrm{R_\odot}$  &
		$1.059~\mathrm{R_{Jup}}$   &   $80.19^\circ$   &  0.268 &\\ \hline
$90^\circ$  &   $0^\circ$  &  $0.0~\mathrm{R_{pole}}$  & $2.30~\mathrm{R_\odot}$  &
		$1.17~\mathrm{R_{Jup}}$   &   $80.30^\circ$   &  0.916  & 1.000 * \\
$90^\circ$  &   $0^\circ$  &  $0.3~\mathrm{R_{pole}}$  & $2.06~\mathrm{R_\odot}$  &
		$1.00~\mathrm{R_{Jup}}$   &   $84.63^\circ$   &  1.000  & 1.000 * \\ 
$90^\circ$  &   $0^\circ$  &  $0.6~\mathrm{R_{pole}}$  & $2.05~\mathrm{R_\odot}$  &
		$1.01~\mathrm{R_{Jup}}$   &   $87.10^\circ$   &  1.000  & 1.000 * \\
$90^\circ$  &   $0^\circ$  &  $0.9~\mathrm{R_{pole}}$  & $2.04~\mathrm{R_\odot}$  &
		$1.01~\mathrm{R_{Jup}}$   &   $89.47^\circ$   &  1.000  & 1.000 * \\\hline
$0^\circ$   &  $90^\circ$  &  $0.0~\mathrm{R_{pole}}$  & $1.89~\mathrm{R_\odot}$  &
		$1.12~\mathrm{R_{Jup}}$   &   $84.57^\circ$   & -1.000  &-1.000 * \\
$0^\circ$   &  $90^\circ$  &  $0.3~\mathrm{R_{pole}}$  & $1.91~\mathrm{R_\odot}$  &
		$1.11~\mathrm{R_{Jup}}$   &   $84.00^\circ$   & -1.000  &-1.000 * \\
$0^\circ$   &  $90^\circ$  &  $0.6~\mathrm{R_{pole}}$  & $1.92~\mathrm{R_\odot}$  &
		$1.04~\mathrm{R_{Jup}}$   &   $82.94^\circ$   & -1.000  &-1.000 * \\
$0^\circ$   &  $90^\circ$  &  $0.9~\mathrm{R_{pole}}$  & $1.85~\mathrm{R_\odot}$  &
		$0.91~\mathrm{R_{Jup}}$   &   $82.12^\circ$   & -0.585  &-1.000 * \\
\hline
\enddata
\tablecomments{Tabled here are the best-fit parameters $R_*$, $R_p$,
$\inclination$, $c_1$, and $c_2$ for fits assuming spherical stars of synthetic 
lightcurves of hypothetical Altair-planet systems with $R_p=\mathrm{R_{Jup}}$
and the planet at $0.05~\mathrm{AU}$.  The stellar obliquity to the plane of the
sky $\obliq$ for each synthetic curve is listed at left, followed by the angle
$\azimuth$
between the planet's orbit pole and the stellar rotation axis projected into the
plane of the sky.  When fixing $c_2$ at zero and fitting only for $c_1$ resulted
in a high-quality fit to the data, $c_2$ is not listed.  Fits that have a `*' next
to their value for $c_2$ did not produce good fits even with the second limb
darkening parameter.}
\end{deluxetable}

I show synthetic transit lightcurves of planets in spin-orbit aligned geometries in
Figure \ref{figure:q0.a0}, for various transit impact parameters $b$.  The lightcurves
are symmetrical.  Furthermore, since the parts of the star that the planet transit chord
passes over all have the same temperature because they equidistant from the stellar
rotation pole, the transit bottom shows normal limb-darkening curvature.  As a result,
the specific lightcurve shapes for spin-orbit aligned transiting planets are
indistinguishable from those for planets orbiting spherical, slow-rotating stars.

The transit depths and durations are different, though.  In particular, the transit depth
\emph{increases} with impact parameter, opposite of the case for spherical stars.  If you
were to fit these lightcurves assuming a spherical star, then the resulting best-fit transit
parameters would be different from the actual ones.  In order to show this effect, I fit the
synthetic transit lightcurves show in Figure \ref{figure:q0.a0} using a Leavenberg-Marquardt
chi-squared minimization scheme from \citet{1992nrca.book.....P} as described in
\citet{oblateness.2003} and show the resulting best-fit parameters in Table~\ref{table:fits}.

For planets transiting across the center of the star ($b=0.0$), the best-fit parameters
retain easily-interpreted astrophysical meaning.  The ratio of radii $R_p/R_*$ comes from the
total transit depth, which in this case is nearly identical to that for the case of a $b=0.0$
transit across a spherical star with radius equal to the equatorial radius of the
fast-rotating star.  In that case, while using the Altair parameters that I use here, the
total larger projected area of the stellar disk makes up for lower flux coming from the polar
regions, yielding similar net stellar flux.  The fitting algorithm thus gets $R_p/R_*$
correct, assuming $R_* = R_{eq}$ and not the real average radius of the projected disk,
$(1-\frac{f}{2})R_{eq}$.  The total duration of the transit and the duration of
ingress and egress fix the impact parameter, which the fit correctly
determines to be near 0.0 ($\inclination \sim 90^\circ$), and the stellar radius, which
is very close to the star's true equatorial radius.  Hence the planetary radius also comes
out correctly.  Limb darkening is right because of the same-temperature effect described
above.

As I consider higher and higher impact parameters, shown in Figure \ref{figure:q0.a0} as
fractions of Altair's polar radius $1.63~\mathrm{R_\odot}$, the best-fit parameters deviate
further and further from the actual values.  As the planet transits across hotter,
brighter parts of the stellar photosphere near the pole, the best-fit value for planet
radius increases by up to 30\% above the input planet radius.  The stellar radius,
driven by the somewhat longer transit durations, drops slightly but stays closer to the
input star's equatorial radius than it does to the true average radius of the stellar
disk.  Less importantly, the inclination and limb darkening parameter are somewhat
underestimated.

The end result of these calculations is that if the planets orbiting fast-rotating stars
are spin-orbit aligned, it may not be evident at first glance.  The transit lightcurves
will not stick out.  Instead the measured transit parameters' errors will be systematic
in nature, and the fits will still be good.  

If the spin-orbit aligned planet case turns out to be prevalent around fast-rotating
stars in the \emph{Kepler} sample, then measuring the spin-orbit alignment for planets
will require a separate identification of the star's fast-rotating status. 
\emph{Kepler}'s photometric precision should be good enough to identify the star's
rotation rate in long timeseries photometry.  If the planets transiting those stars show
normal-looking lightcurves, that would indicate that the planet is spin-orbit aligned. 
If most or all planets are spin-orbit aligned, then they likely either formed in-situ or
migrated there, and were not scattered by close encounters with other planets.

\section{SPIN-ORBIT MISALIGNED}

\begin{figure*}
\epsscale{2.1}
\plotone{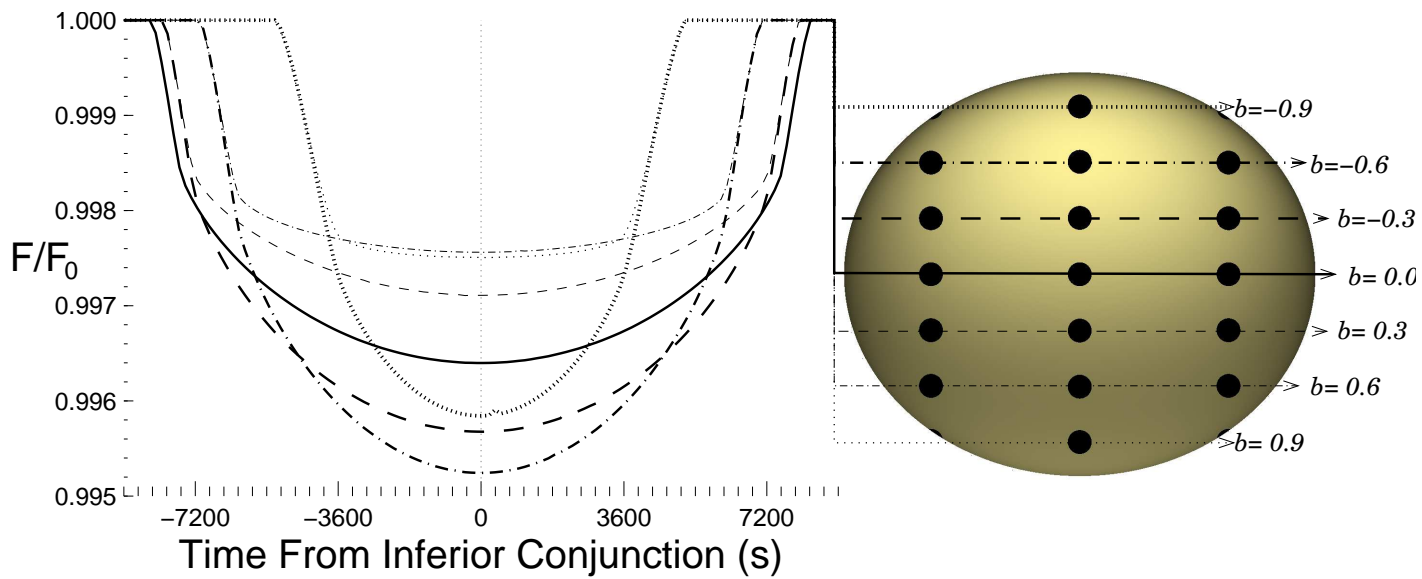}
\caption{Synthetic lightcurves for transiting 1 $R_{Jup}$ in a $0.05~\mathrm{AU}$
orbit around an Altair-like star with obliquity $30^\circ$ are plotted.  The seven
curves  correspond to planets with
transit impact parameters of $b=-0.9~R_{pole}$ (thick dotted), $b=-0.6~R_{pole}$
(thick dot-dashed), $b=-0.3~R_{pole}$ (thick dashed), $b=0.0~R_{pole}$ (thick solid), $b=0.3~R_{pole}$ (dashed),
$b=0.6~R_{pole}$ (dot-dashed), and $b=0.9~R_{pole}$ (dotted).  The curves' shapes are
start to differ from those of transits of slow-rotating, spherical stars. 
The ingress and egress of transits at opposite impact parameter
are nearly the same.  The transit bottoms differ.  Transits near the hot stellar 
north pole are deep and show severe curvature of the
transit bottom.
\label{figure:q30.a0}}
\end{figure*}

\begin{figure*}
\epsscale{2.1}
\plotone{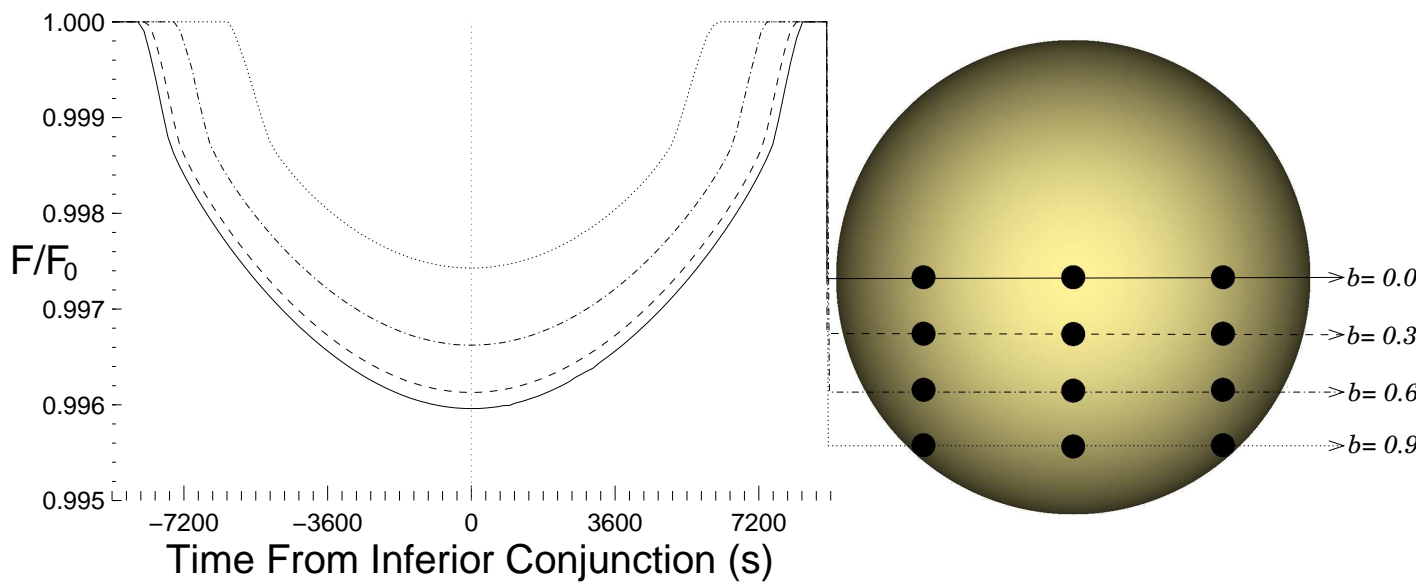}
\caption{ Synthetic lightcurves for transiting 1 $R_{Jup}$ in a $0.05~\mathrm{AU}$
orbit around an Altair-like star with obliquity $90^\circ$ (pole-on) are plotted. 
The four
curves  correspond to planets with
transit impact parameters of $b=0.0~R_{pole}$ (solid), $b=0.3~R_{pole}$ (dashed),
$b=0.6~R_{pole}$ (dot-dashed), and $b=0.9~R_{pole}$ (dotted).  The lightcurves
show a pronounced `U'-shape that is typically considered to be characteristic of
grazing eclipsing binary stars rather than planets.  The extreme curvature of the
usually flat planet transit bottom makes the locations of second and third contact
difficult to discern.  Due to symmetry around the stellar pole, all transits with
these same impact parameters will look the same regardless of $\azimuth$.
\label{figure:q90.a0}}
\end{figure*}

\begin{figure*}
\epsscale{2.1}
\plotone{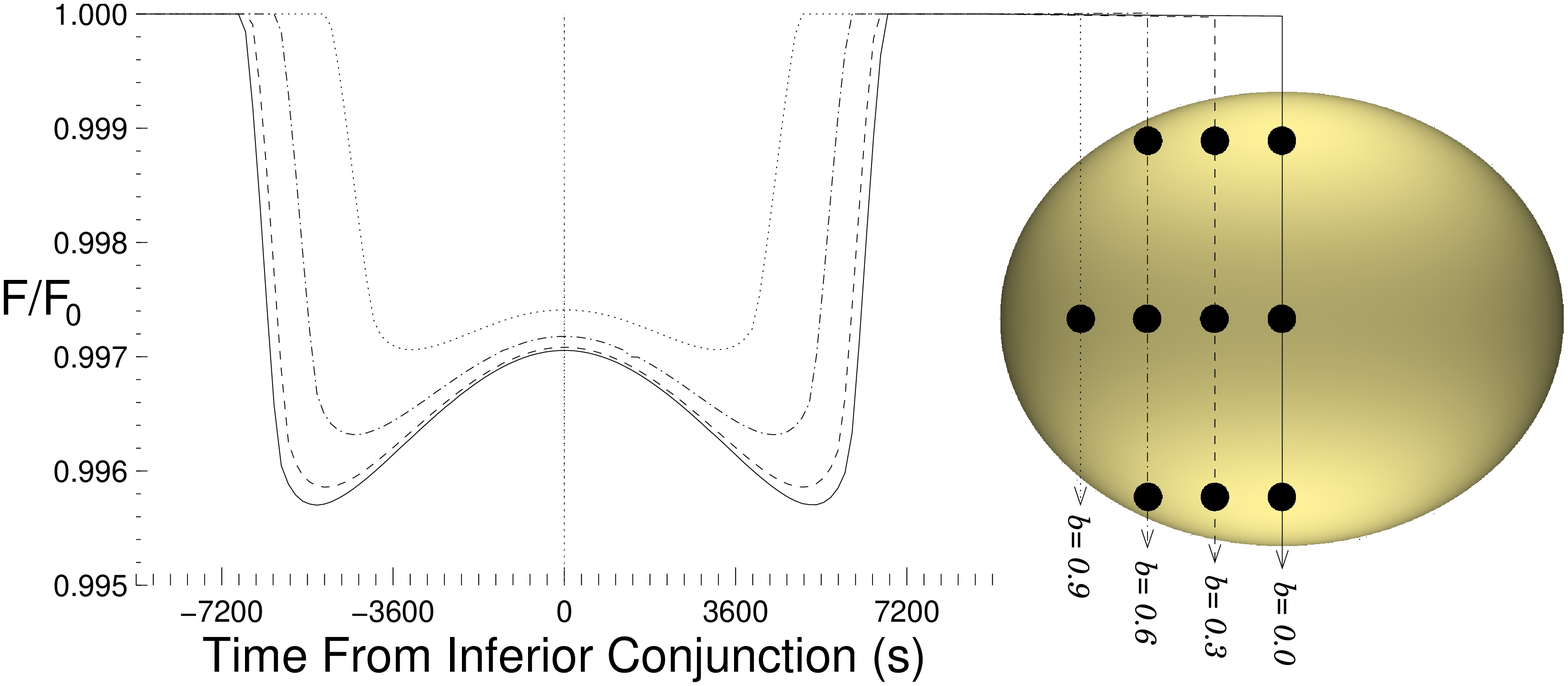}
\caption{
Synthetic lightcurves for transiting 1 $\mathrm{R_{Jup}}$ in a 
$0.05~\mathrm{AU}$
orbit around an Altair-like star with obliquity $0^\circ$ (equator-on) are plotted,
similar to Figure \ref{figure:q0.a0}, but this time with an azimuthal angle of 
$\azimuth=90^\circ$.  The four curves  correspond to planets with
transit impact parameters of $b=0.0~R_{pole}$ (solid), $b=0.3~R_{pole}$ (dashed),
$b=0.6~R_{pole}$ (dot-dashed), and $b=0.9~R_{pole}$ (dotted).  This unlikely
$90^\circ$ transit geometry also produces symmetric transit lightcurves, albeit
highly unusual ones.  These curves are deepest near second and third contact, and
shallow at mid-transit.  If assuming a spherical-star model, then the
interpretation of this ``double-horned" structure might be a negative
limb-darkening coefficient possibly associated with a temperature inversion in the
stellar atmosphere; however, such a model fits these data very poorly (see text).
\label{figure:q0.a90} }
\end{figure*}

If instead the transiting planets around fast-rotating stars experienced close
encounters that threw them out of spin-aligned orbits, then more exotic lightcurves
result.  Recent Rossiter-McLaughlin results from transiting planetary systems indicate
that this situation may be more common than previously thought
\citep{2009IAUS..253..508H}.  Given the difficulty of radial velocity, and thus
Rossiter-McLaughlin, measurements around early-type stars, transit photometry may then
be the best way to measure the spin-orbit alignment in those systems.

\subsection{SYMMETRIC}

Just because the transit lightcurve of a planet around a fast-rotating star is symmetric
does not mean that the planet is spin-orbit aligned.  Any planet whose transit chord is
perpendicular to the projected stellar rotation pole, \emph{i.e.} for which the angle
between the transit chord and the projected pole, $\azimuth$, is $0.0^\circ$, will show
a symmetric lightcurve.  I plot the lightcurve for a few such hypothetical planets in
Figures \ref{figure:q30.a0}  and \ref{figure:q90.a0}.  Figure \ref{figure:q30.a0} shows
planets around a star with obliquity $\obliq=30^\circ$.  Figure \ref{figure:q90.a0}
shows transits for a star viewed pole-on with $\obliq=90^\circ$.

Although these lightcurves are symmetrical about the mid-transit point, unlike the
spin-aligned case they are not well-fit using a model that assumes a spherical star.  In
the $\obliq=30^\circ$ case the stellar obliquity removes the symmetry in the stellar disk,
thereby making transits toward the north and south poles distinct from one another.  I
arbitrarily choose to define those transits toward the north of the center of the stellar
disk to have a negative impact parameter so as to differentiate the two possible cases.

In the positive impact parameter cases in Figure \ref{figure:q30.a0}, the planet crosses the
cooler equatorial regions of the star.  Hence those transits are relatively shallow.  The
geometry of the orthographic projection of the star as seen from Earth means that for the high
impact parameter, $b=0.6~R_{pole}$ and $b=0.9~R_{pole}$ cases, the planet covers more southerly
stellar latitudes at ingress and egress than it does at mid-transit.  Because those more
southerly latitudes are hotter, the net result is a flatter lightcurve bottom than occurs for
spherical stars with the same limb darkening parameter ($c_1=0.640$).  The flatness appears as
a lower limb darkening parameter $c_1$ in Table \ref{table:fits}.  For the $b=0.6~R_{pole}$
and $b=0.9~R_{pole}$ cases the model overestimates the star's radius; the best-fit radius
is greater than even the star's equatorial radius.  

For $b=0.0~R_{pole}$ and $b=0.3~R_{pole}$ in the $\obliq=30^\circ$ case, the planet transits
closer to the star's north pole, and hence the transit depth becomes progressively greater. 
The ingress and egress are again at more southerly latitudes than mid-transit.  But now
mid-transit is north of the equator, so the ingress and egress are cooler than mid-transit,
enhancing the curvature of the lightcurve bottom.  I was unable to find satisfactory fits to
either of these synthetic lightcurves using just the first \citet{2001ApJ...552..699B} limb
darkening coefficient $c_1$ alone, even though it was the only one used to generate the
synthetic lightcurve.  However, by allowing the fitting algorithm to optimize $c_2$ as well I
managed to find reasonable fits for $b=0.0~R_{pole}$ and $b=0.3~R_{pole}$.

When fitting $b=-0.3~R_{pole}$, $b=-0.6~R_{pole}$, and $b=-0.9~R_{pole}$ not even
two-limb-darkening fits were acceptable.  The curvature on these transits is so extreme
that  even though they are symmetric, they are discernable from reasonable
spherical-star models  because of the severity of their curvature.  In fact, the
curvature of these planets' transit bottoms are so severe that the lightcurves look
`V'-shaped instead of `U'-shaped.  The transit bottom is difficult to discern.  The
resulting rounded shape might be mistaken for an eclipsing binary star.

Fits for transits of pole-on oriented fast-rotating stars are similar.  As shown in Figure
\ref{figure:q90.a0}, the center of the stellar disk is particularly bright relative to the
limb owing to the center being the hot pole and the limb being the cool equator.  Thus here
again the curvature of the transit lightcurve between second and third contacts is large, and
cannot be satisfactorily fit by a spherical-star model.  In both these cases the difficulty in
identifying these type of planetary transits in the \emph{Kepler} data will probably be
recognizing that they are not eclipsing binaries, not in thinking that they are planets around
spherical stars.

An observer would also measure a symmetrical transit lightcurve if a planet orbits a zero
obliquity ($\obliq=0.0^\circ$) star such that its orbit plane contains the stellar rotation
pole, \emph{i.e.} with $\azimuth=90^\circ$.  I show synthetic lightcurves for such a situation
in Figure \ref{figure:q0.a90}.  These are strange transits.  By moving parallel to the star's
orbit pole, the planets first encounter the limb-darkened edge of the star.  Shortly after
second contact, they cover the brightest part of the stellar photosphere underneath the
transit chord.  The lightcurve is deepest there.  Over the equator at mid-transit the transit
depth is shallow, and then the process repeats itself backwards on egress.

The resulting lightcurve has a ``double-horned" structure that bears a resemblance to what a
transit across a limb-brightened star would look like.  However, spherical-star model fits
cannot reproduce the specific structure.  In particular, a limb-brightened star would have
sharp points at second and third contacts where the derivitive is discontinuous and changes
sign.  The fast-rotating star $\obliq=0^\circ$ $\azimuth=90^\circ$ transits in Figure
\ref{figure:q0.a90} instead have rounded horns at second and third contact owing to the
combination of limb- and gravity-darkening.  The resulting rounded double-horn shape is both
characteristic and diagnostic of this kind of transit.

\subsection{ASYMMETRIC}

\begin{figure*}
\epsscale{2.1}
\plotone{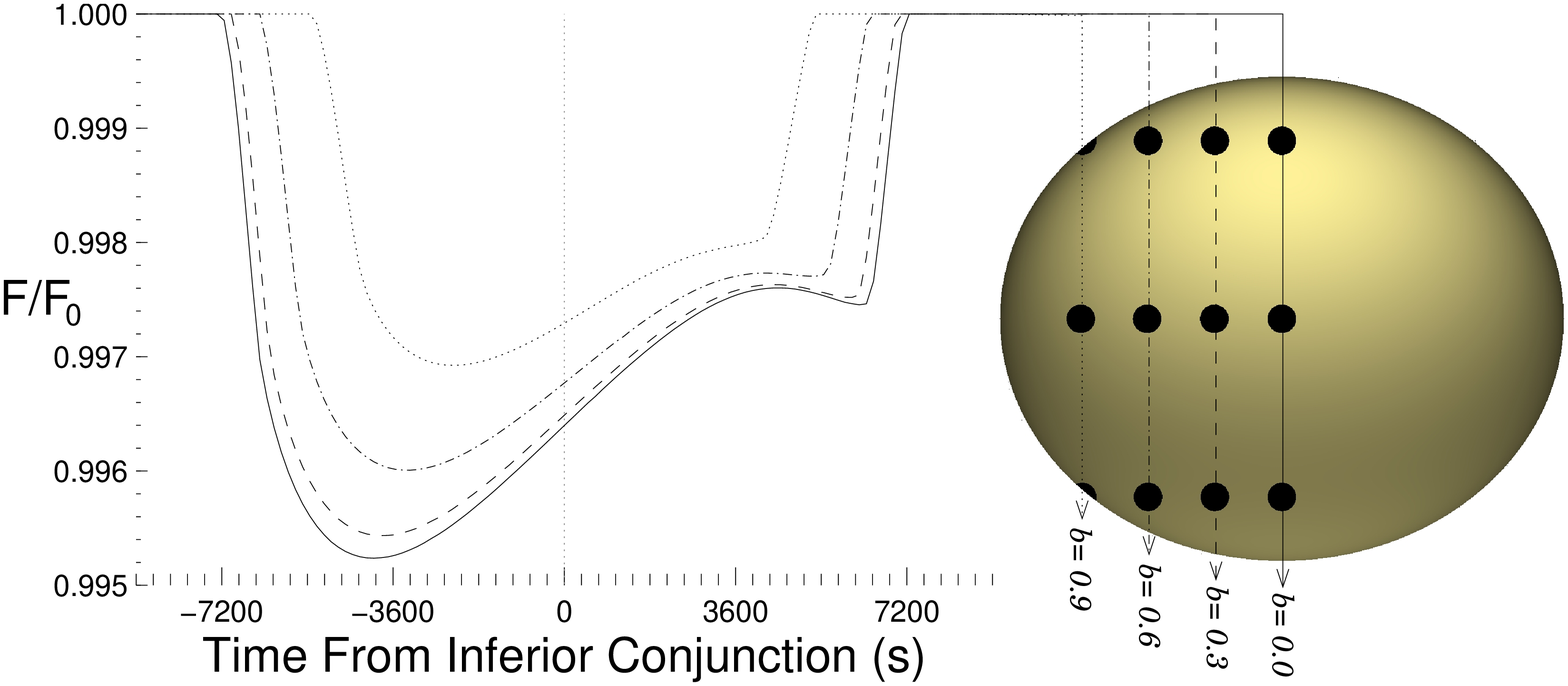}
\caption{ 
Synthetic lightcurves for transiting 1 $\mathrm{R_{Jup}}$ in a 
$0.05~\mathrm{AU}$
orbit around an Altair-like star with obliquity $30^\circ$ are plotted,
similar to Figure \ref{figure:q30.a0}, but this time with an azimuthal angle of 
$\azimuth=90^\circ$.  The four curves  correspond to planets with
transit impact parameters of $b=0.0~R_{pole}$ (solid), $b=0.3~R_{pole}$ (dashed),
$b=0.6~R_{pole}$ (dot-dashed), and $b=0.9~R_{pole}$ (dotted).  The resulting lightcurves are
highly asymmetric, being deeper on the side of the lightcurve where the planet passes over
the hot northern stellar pole.
\label{figure:q30.a90}}
\end{figure*}


\begin{figure*}
\epsscale{2.1}
\plotone{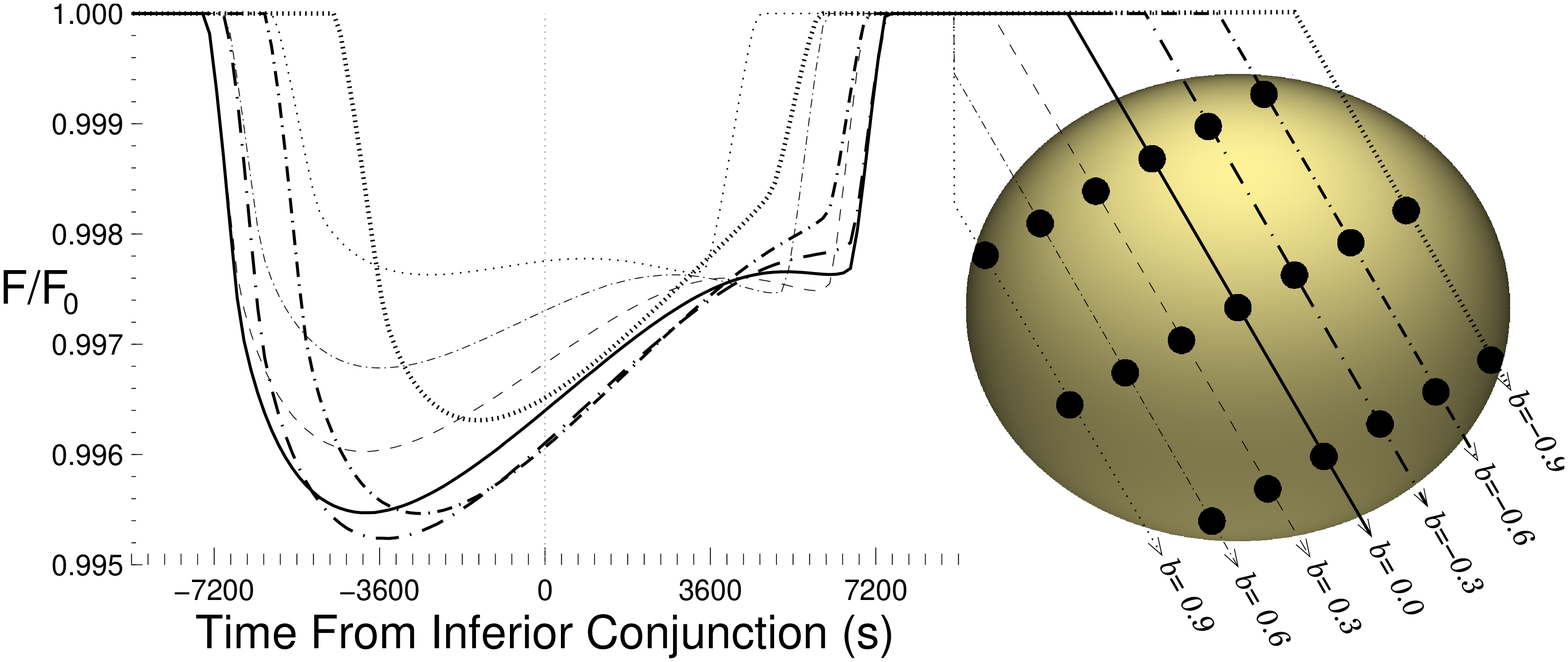}
\caption{ Synthetic lightcurves for transiting 1 $\mathrm{R_{Jup}}$ in a  $0.05~\mathrm{AU}$
orbit around an Altair-like star with obliquity $30^\circ$ are plotted, similar to Figure
\ref{figure:q30.a0} but an azimuthal angle of  $\azimuth=60^\circ$.  The seven curves 
correspond to planets with transit impact parameters of $b=-0.9~R_{pole}$ (thick dotted),
$b=-0.6~R_{pole}$ (thick dot-dashed), $b=-0.3~R_{pole}$ (thick dashed), $b=0.0~R_{pole}$
(thick solid), $b=0.3~R_{pole}$ (dashed), $b=0.6~R_{pole}$ (dot-dashed), and $b=0.9~R_{pole}$
(dotted).  With this oblique azimuth, the photometric lightcurve center does not correspond
with the point where the planet is nearest to Earth.  The lightcurves show a diversity of
complex asymmetric shapes as a function of impact parameter.
\label{figure:q30.a60}}
\end{figure*}


\begin{figure*}
\epsscale{2.1}
\plotone{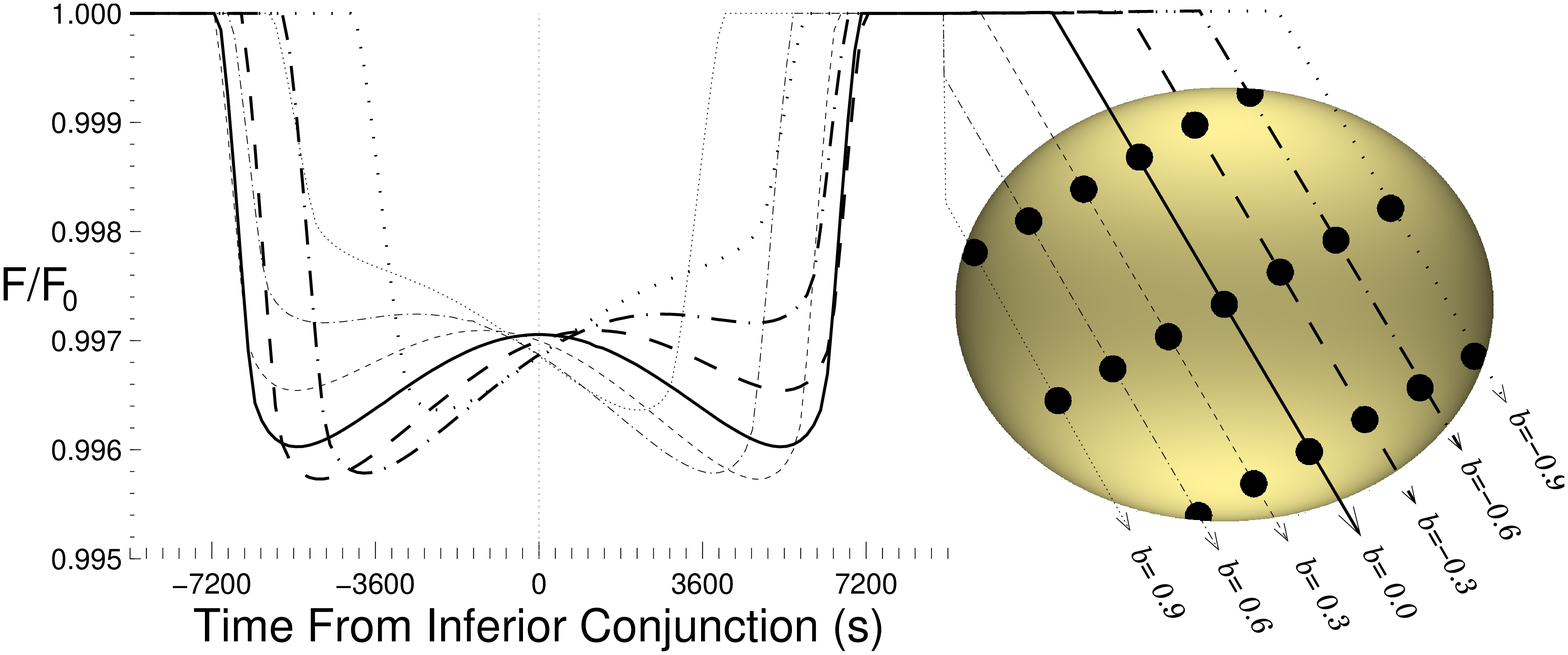}
\caption{Synthetic lightcurves for transiting 1 $\mathrm{R_{Jup}}$ in a  $0.05~\mathrm{AU}$
orbit around an Altair-like star are plotted, similar to Figure
\ref{figure:q30.a60} but with an obliquity $\obliq=0^\circ$.  The seven curves 
correspond to planets with transit impact parameters of $b=-0.9~R_{pole}$ (thick dotted),
$b=-0.6~R_{pole}$ (thick dot-dashed), $b=-0.3~R_{pole}$ (thick dashed), $b=0.0~R_{pole}$
(thick solid), $b=0.3~R_{pole}$ (dashed), $b=0.6~R_{pole}$ (dot-dashed), and $b=0.9~R_{pole}$
(dotted).  Due to the symmetry of the stellar disk in this case the $b=0.0~R_{pole}$ case
shows a symmetric lightcurve, and positive impact parameters have time-reversed
lightcurves when compared to their negative counterparts.
\label{figure:q0.a60}}
\end{figure*}

If the stellar orientation and transit chord have more exotic geometries, then highly unusual
asymmetric lightcurves result.  A star with obliquity $\obliq=30^\circ$ and transit chord
azimuth $\azimuth=90^\circ$ (motion parallel to the projected stellar rotation axis) presents
an easy such case to visualize (Figure \ref{figure:q30.a90}).  Because the stellar disk
presents only a left-right symmetry, these transits pass over very different photospheric
temperatures in the first half of the transit relative to the second half.  

In the $\obliq=30^\circ$, $\azimuth=90^\circ$, impact parameter $b=0.0~\mathrm{R_{pole}}$
case, the planet first crosses a darkened stellar limb before occulting the star's hot north
pole.  The resulting lightcurve is deepest at this point.  After mid-transit the planet
occults the cool equator for a shallower transit depth, but then the depth increases again
near third contact as the planet covers higher southern latitudes near the southern stellar
limb.  The same process occurs in a more abbreviated fashion as the impact parameter $b$
increases.

If the transit chord azimuth is oblique, then the lightcurves can become quite complex as
shown in Figure \ref{figure:q30.a60}.  The central $b=0.0~\mathrm{R_{pole}}$ transit resembles
those from Figure \ref{figure:q30.a90}.  The more northerly transits (those defined to have
negative impact parameters) also show strong lightcurve asymmetries, with deep first halves
and shallowing second halves.  Some of the lightcurves turn over again near third contact, but
the ones with more negative impact parameters do not.

For postitive, more southerly impact parameters, the depth asymmetry decreases.  Futher from
the hot pole the temperatures under the transit chord are more uniform.  The
$b=0.9~\mathrm{R_{pole}}$ transit is particularly interesting.  It shows a nearly uniform
depth in time, but while the ingress is long and gently curving, the egress is abrupt.  A
higher photometric precision would be required to definitively identify such a transit
lightcurve as being one from a fast-rotating star than would be necessary for some of the more
spectacularly asymmetric lightcurves.

With oblique azimuths also come uncentered lightcurves.  Because of the stellar oblateness,
the center of these transits does not occur at the time of the planet's inferior conjunction, 
that is,
when the planet is closest to the Earth.  The total discrepancy can be up to a few tenths of
the total transit duration, depending on the stellar oblateness and the transit geometry.

The oblique azimuth transits of fast-rotating stars also introduce an asymmetry in the
duration of transit ingress and egress.  This effect comes about because the angle between the
stellar limb and the transit chord is different on ingress and egress, thereby causing the
transiting planet to take different amounts of time to cross the limb in each case.

Lightcurve asymmetries can occur in zero-obliquity ($\obliq=0^\circ$) stars as well when the
transit chord azimuth is oblique.  I show such a case in Figure \ref{figure:q0.a60}, with
$\obliq=0^\circ$ and $\azimuth=60^\circ$.  The central $b=0.0~\mathrm{R_{pole}}$ transit
is symmetric, similar to those from Figure \ref{figure:q0.a90}.  Non-central transits are
asymmetric and resemble those from Figure \ref{figure:q30.a90} with one half of the transit
being deeper than the other.  In this $\obliq=0^\circ$ oblique case, the transits with
negative impact parameter are the time-inverse of those with positive impact parameter.





\section{COLOR EFFECTS}
 
\begin{figure*}
\epsscale{2.1}
\plotone{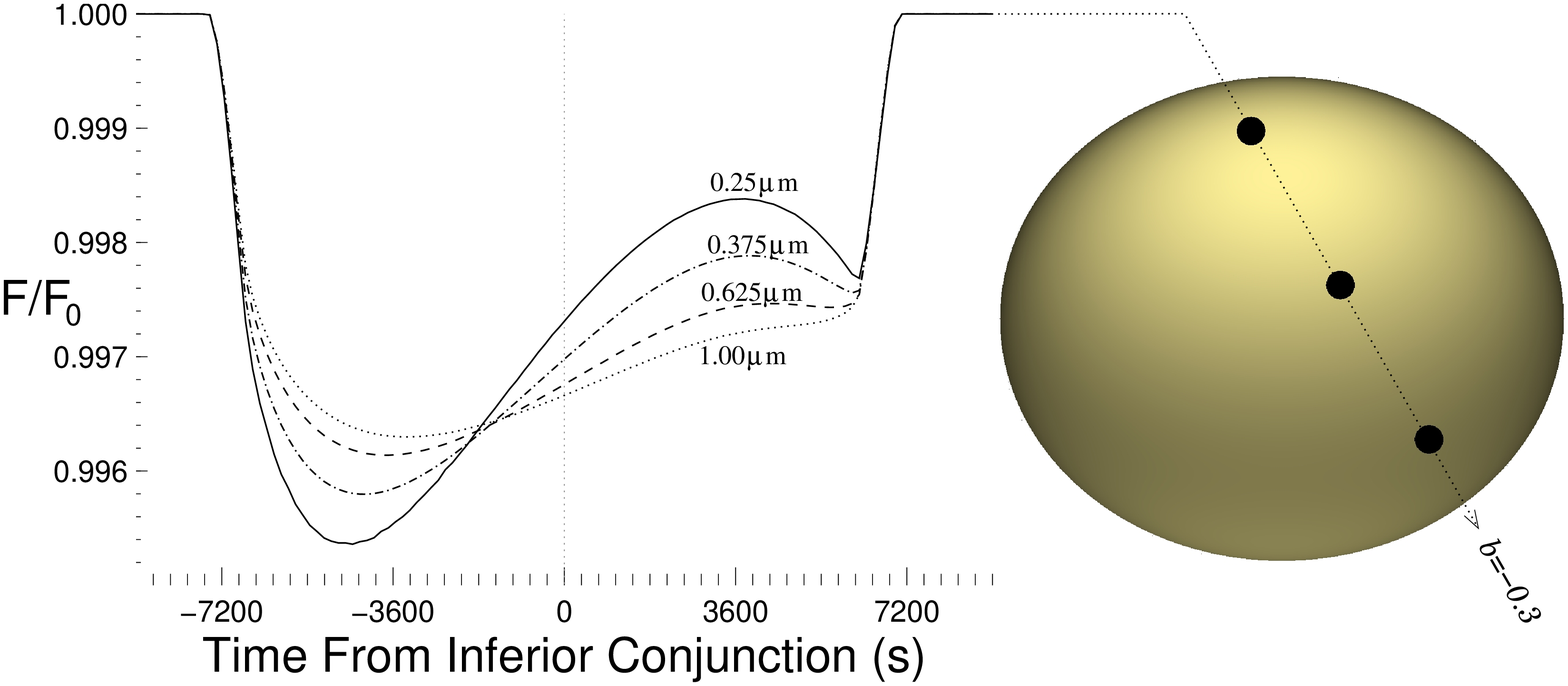}
\caption{Synthetic lightcurves for transiting 1 $\mathrm{R_{Jup}}$ in a $0.05~\mathrm{AU}$
orbit around an Altair-like star with $\obliq=60^\circ$ are plotted.  The 4 curves all
correspond to planets with a transit impact parameter of $b=-0.3~R_{pole}$.  The
different curves correspond to what a lightcurve would look like if it were acquired at
different wavelenghts:  $0.25~\mathrm{\mu m}$ (solid line), $0.375~\mathrm{\mu m}$ 
(dotted line).  Note that wavelength is all that is changed here -- in particular I have
employed the identical limb darkening parameter of $c_1=0.64$ for each wavelength. 
Though this is unphysical in that the limb darkening should diminish with increasing
wavelength, this way makes it easiest to identify just the differences due to wavelength
changes.  The contrast on the stellar disk is greater on the Wien (short wavelength) 
end of the blackbody
curve, while the contrast is muted on the Rayleigh-Jeans (long wavelength) side.
\label{figure:colors}}
\end{figure*}

Due to the temperature nonuniformity across the stellar disk, transit lightcurves around
rapidly-rotating stars differ as a function of wavelength.  I show an example of
this effect in Figure \ref{figure:colors}.  While a hotter blackbody radiates more flux
at all wavelengths than a colder one, the flux ratio (hotter over colder) is greater at
the Wien end of the blackbody curve.  Hence the effects of rapid rotation in transit
lightcurves are maximized at short wavelengths.

At wavelengths much longer than the blackbody emission peak, the flux ratio between the
hotter and colder areas approaches the temperature ratio in Kelvin.  Utilizing multiple
wavelength lightcurve observations would strongly constrain the temperature structure on
the stellar disk in addition to the transit parameters.  This effect will be important
when fitting for a fully-consistent model for the star (instead of assuming Altair's
parameters as I do in this paper).

\section{CONCLUSION}

The gravity-darkening effect for rapidly rotating, oblate stars first predicted by
\citet{1924MNRAS..84..665V} allows for highly unusual lightcurves when such a star is
transited by an extrasolar planet.  The distinctive lightcurves allow for a solely
photometric determination of the relative alignment between the stellar spin axis and the
planet's orbit normal (spin-orbit alignment), a measurement that usually requires radial
velocity measurements of the Rossiter-McLaughlin effect.  The alignment bears a
fingerprint of the planet's formation and evolution.  Planets that form in an orderly
fashion and migrate within a disk ought to end up coplanar with the stellar equator. 
Those that have experienced planet-planet scattering events ought in general to not be
spin-orbit aligned.

Spin-orbit aligned planets around fast-rotating stars show lightcurves that resemble
those of planets orbiting slow-rotating stars so closely that the two cannot be
distinguished based on lightcurves alone.  However, fitting the lightcurves of the
planets around fast-rotators with a spherical-star model yields incorrect transit
parameters.  The significant systematic errors introduced into the determination of the
planet's radius are particularly distressing, especially at high impact parameters. 
This degeneracy can be resolved by independent  determination of the stellar rotation
period.

Planets that are not spin-orbit aligned can lead to spectacularly strange lightcurves
that might not otherwise be immediately recognizable as planetary transits.  Such
planets can yield asymmetric lightcurves and highly curved transit floors that rule out
the spherical-star hypothesis.  The precise shape of such transits reveals the stellar
obliquity and the angle between the transit chord and the projected stellar rotation
axis.  Together these two values determine the net spin-orbit alignment of the system. 
There is still a twofold degeneracy of the system, reflected around the projected
stellar rotation axis.  

The changing temperatures across the face of the stellar disk lead to strong variability
of transit lightcurves as a function of wavelength.  Multi-wavelength transit photometry
can then directly constrain the stellar temperatures, allowing for better models of the
stars themselves.  High-quality stellar models lead to more reliable measurements of the
planet's radius.

Perhaps $\sim5-10\%$ of the \emph{Kepler} target stars ought to be main sequence dwarfs
of spectral type early F or earlier, and thus are probably rapid rotators.  Determination
of the spin-orbit alignment of planets orbiting these stars will provide a glimpse into
the planet formation process around high-mass stars for the first time.

\acknowledgements

\bibliographystyle{apj}
\bibliography{references}

\begin{thebibliography}{37}
\expandafter\ifx\csname natexlab\endcsname\relax\def\natexlab#1{#1}\fi

\bibitem[{{Barnes}(2007)}]{2007PASP..119..986B}
{Barnes}, J.~W. 2007, \pasp, 119, 986

\bibitem[{{Barnes} \& {Fortney}(2003)}]{oblateness.2003}
{Barnes}, J.~W. \& {Fortney}, J.~J. 2003, \apj, 588, 545

\bibitem[{{Barnes} \& {Fortney}(2004)}]{2004ApJ...616.1193B}
---. 2004, \apj, 616, 1193

\bibitem[{{Brown} {et~al.}(2001){Brown}, {Charbonneau}, {Gilliland}, {Noyes},
  \& {Burrows}}]{2001ApJ...552..699B}
{Brown}, T.~M., {Charbonneau}, D., {Gilliland}, R.~L., {Noyes}, R.~W., \&
  {Burrows}, A. 2001, \apj, 552, 699

\bibitem[{{Ciardi} {et~al.}(2001){Ciardi}, {van Belle}, {Akeson}, {Thompson},
  {Lada}, \& {Howell}}]{2001ApJ...559.1147C}
{Ciardi}, D.~R., {van Belle}, G.~T., {Akeson}, R.~L., {Thompson}, R.~R.,
  {Lada}, E.~A., \& {Howell}, S.~B. 2001, \apj, 559, 1147

\bibitem[{{Djura{\v s}evi{\'c}} {et~al.}(2003){Djura{\v s}evi{\'c}},
  {Rovithis-Livaniou}, {Rovithis}, {Georgiades}, {Erkapi{\'c}}, \&
  {Pavlovi{\'c}}}]{2003A&A...402..667D}
{Djura{\v s}evi{\'c}}, G., {Rovithis-Livaniou}, H., {Rovithis}, P.,
  {Georgiades}, N., {Erkapi{\'c}}, S., \& {Pavlovi{\'c}}, R. 2003, \aap, 402,
  667

\bibitem[{{Hansen} {et~al.}(2004){Hansen}, {Kawaler}, \&
  {Trimble}}]{2004sipp.book.....H}
{Hansen}, C.~J., {Kawaler}, S.~D., \& {Trimble}, V. 2004, {Stellar interiors :
  physical principles, structure, and evolution} (Stellar interiors : physical
  principles, structure, and evolution, 2nd ed., by C.J.~Hansen, S.D.~Kawaler,
  and V.~Trimble.~New York: Springer-Verlag, 2004.)

\bibitem[{{H{\'e}brard} {et~al.}(2009){H{\'e}brard}, {Bouchy}, {Pont},
  {Loeillet}, {Rabus}, {Bonfils}, {Moutou}, {Boisse}, {Delfosse}, {Desort},
  {Eggenberger}, {Ehrenreich}, {Forveille}, {Lagrange}, {Lovis}, {Mayor},
  {Pepe}, {Perrier}, {Santos}, {Queloz}, {S{\'e}gransan}, {Udry}, \&
  {Vidal-Madjar}}]{2009IAUS..253..508H}
{H{\'e}brard}, G., {Bouchy}, F., {Pont}, F., {Loeillet}, B., {Rabus}, M.,
  {Bonfils}, X., {Moutou}, C., {Boisse}, I., {Delfosse}, X., {Desort}, M.,
  {Eggenberger}, A., {Ehrenreich}, D., {Forveille}, T., {Lagrange}, A.-M.,
  {Lovis}, C., {Mayor}, M., {Pepe}, F., {Perrier}, C., {Santos}, N.~C.,
  {Queloz}, D., {S{\'e}gransan}, D., {Udry}, S., \& {Vidal-Madjar}, A. 2009, in
  IAU Symposium, Vol. 253, IAU Symposium, 508--511

\bibitem[{{Johnson} {et~al.}(2007){Johnson}, {Fischer}, {Marcy}, {Wright},
  {Driscoll}, {Butler}, {Hekker}, {Reffert}, \& {Vogt}}]{2007ApJ...665..785J}
{Johnson}, J.~A., {Fischer}, D.~A., {Marcy}, G.~W., {Wright}, J.~T.,
  {Driscoll}, P., {Butler}, R.~P., {Hekker}, S., {Reffert}, S., \& {Vogt},
  S.~S. 2007, \apj, 665, 785

\bibitem[{{Johnson} {et~al.}(2008{\natexlab{a}}){Johnson}, {Marcy}, {Fischer},
  {Wright}, {Reffert}, {Kregenow}, {Williams}, \& {Peek}}]{2008ApJ...675..784J}
{Johnson}, J.~A., {Marcy}, G.~W., {Fischer}, D.~A., {Wright}, J.~T., {Reffert},
  S., {Kregenow}, J.~M., {Williams}, P.~K.~G., \& {Peek}, K.~M.~G.
  2008{\natexlab{a}}, \apj, 675, 784

\bibitem[{{Johnson} {et~al.}(2008{\natexlab{b}}){Johnson}, {Winn}, {Narita},
  {Enya}, {Williams}, {Marcy}, {Sato}, {Ohta}, {Taruya}, {Suto}, {Turner},
  {Bakos}, {Butler}, {Vogt}, {Aoki}, {Tamura}, {Yamada}, {Yoshii}, \&
  {Hidas}}]{2008ApJ...686..649J}
{Johnson}, J.~A., {Winn}, J.~N., {Narita}, N., {Enya}, K., {Williams},
  P.~K.~G., {Marcy}, G.~W., {Sato}, B., {Ohta}, Y., {Taruya}, A., {Suto}, Y.,
  {Turner}, E.~L., {Bakos}, G., {Butler}, R.~P., {Vogt}, S.~S., {Aoki}, W.,
  {Tamura}, M., {Yamada}, T., {Yoshii}, Y., \& {Hidas}, M. 2008{\natexlab{b}},
  \apj, 686, 649

\bibitem[{{Juri{\'c}} \& {Tremaine}(2008)}]{2008ApJ...686..603J}
{Juri{\'c}}, M. \& {Tremaine}, S. 2008, \apj, 686, 603

\bibitem[{Kalas {et~al.}(2008)Kalas, Graham, Chiang, Fitzgerald, Clampin, Kite,
  Stapelfeldt, Marois, \& Krist}]{PaulKalas11282008}
Kalas, P., Graham, J.~R., Chiang, E., Fitzgerald, M.~P., Clampin, M., Kite,
  E.~S., Stapelfeldt, K., Marois, C., \& Krist, J. 2008, Science, 322, 1345

\bibitem[{{Liu} {et~al.}(2009){Liu}, {Sato}, {Zhao}, \& {Ando}}]{2009LiuRIAA}
{Liu}, Y.-J., {Sato}, B., {Zhao}, G., \& {Ando}, H. 2009, Research in Astronomy
  and Astrophysics, 9, 1

\bibitem[{{Lovis} \& {Mayor}(2007)}]{2007A&A...472..657L}
{Lovis}, C. \& {Mayor}, M. 2007, \aap, 472, 657

\bibitem[{{Maeder}(2009)}]{2009pfer.book.....M}
{Maeder}, A. 2009, {Physics, Formation and Evolution of Rotating Stars}
  (Physics, Formation and Evolution of Rotating Stars: , Astronomy and
  Astrophysics Library, Volume .~ISBN 978-3-540-76948-4.~Springer Berlin
  Heidelberg, 2009)

\bibitem[{Marois {et~al.}(2008)Marois, Macintosh, Barman, Zuckerman, Song,
  Patience, Lafreniere, \& Doyon}]{ChristianMarois11282008}
Marois, C., Macintosh, B., Barman, T., Zuckerman, B., Song, I., Patience, J.,
  Lafreniere, D., \& Doyon, R. 2008, Science, 322, 1348

\bibitem[{{Monnier} {et~al.}(2007){Monnier}, {Zhao}, {Pedretti}, {Thureau},
  {Ireland}, {Muirhead}, {Berger}, {Millan-Gabet}, {Van Belle}, {ten
  Brummelaar}, {McAlister}, {Ridgway}, {Turner}, {Sturmann}, {Sturmann}, \&
  {Berger}}]{2007Sci...317..342M}
{Monnier}, J.~D., {Zhao}, M., {Pedretti}, E., {Thureau}, N., {Ireland}, M.,
  {Muirhead}, P., {Berger}, J.-P., {Millan-Gabet}, R., {Van Belle}, G., {ten
  Brummelaar}, T., {McAlister}, H., {Ridgway}, S., {Turner}, N., {Sturmann},
  L., {Sturmann}, J., \& {Berger}, D. 2007, Science, 317, 342

\bibitem[{{Moutou} {et~al.}(2009){Moutou}, {H{\'e}brard}, {Bouchy},
  {Eggenberger}, {Boisse}, {Bonfils}, {Gravallon}, {Ehrenreich}, {Forveille},
  {Delfosse}, {Desort}, {Lagrange}, {Lovis}, {Mayor}, {Pepe}, {Perrier},
  {Pont}, {Queloz}, {Santos}, {S{\'e}gransan}, {Udry}, \&
  {Vidal-Madjar}}]{2009A&A...498L...5M}
{Moutou}, C., {H{\'e}brard}, G., {Bouchy}, F., {Eggenberger}, A., {Boisse}, I.,
  {Bonfils}, X., {Gravallon}, D., {Ehrenreich}, D., {Forveille}, T.,
  {Delfosse}, X., {Desort}, M., {Lagrange}, A.-M., {Lovis}, C., {Mayor}, M.,
  {Pepe}, F., {Perrier}, C., {Pont}, F., {Queloz}, D., {Santos}, N.~C.,
  {S{\'e}gransan}, D., {Udry}, S., \& {Vidal-Madjar}, A. 2009, \aap, 498, L5

\bibitem[{{Narita} {et~al.}(2009){Narita}, {Hirano}, {Sato}, {Winn}, {Suto},
  {Turner}, {Aoki}, {Tamura}, \& {Yamada}}]{2009arXiv0905.4727N}
{Narita}, N., {Hirano}, T., {Sato}, B., {Winn}, J.~N., {Suto}, Y., {Turner},
  E.~L., {Aoki}, W., {Tamura}, M., \& {Yamada}, T. 2009, ArXiv e-prints

\bibitem[{{Niedzielski} {et~al.}(2008){Niedzielski}, {Go\'zdziewski},
  {Wolszczan}, {Konacki}, {Nowak}, \& {Zieli{\'n}ski}}]{2008arXiv0810.1710N}
{Niedzielski}, A., {Go\'zdziewski}, K., {Wolszczan}, A., {Konacki}, M.,
  {Nowak}, G., \& {Zieli{\'n}ski}, P. 2008, ArXiv e-prints

\bibitem[{{Niedzielski} {et~al.}(2007){Niedzielski}, {Konacki}, {Wolszczan},
  {Nowak}, {Maciejewski}, {Gelino}, {Shao}, {Shetrone}, \&
  {Ramsey}}]{2007ApJ...669.1354N}
{Niedzielski}, A., {Konacki}, M., {Wolszczan}, A., {Nowak}, G., {Maciejewski},
  G., {Gelino}, C.~R., {Shao}, M., {Shetrone}, M., \& {Ramsey}, L.~W. 2007,
  \apj, 669, 1354

\bibitem[{{Peterson} {et~al.}(2006{\natexlab{a}}){Peterson}, {Hummel}, {Pauls},
  {Armstrong}, {Benson}, {Gilbreath}, {Hindsley}, {Hutter}, {Johnston},
  {Mozurkewich}, \& {Schmitt}}]{2006ApJ...636.1087P}
{Peterson}, D.~M., {Hummel}, C.~A., {Pauls}, T.~A., {Armstrong}, J.~T.,
  {Benson}, J.~A., {Gilbreath}, G.~C., {Hindsley}, R.~B., {Hutter}, D.~J.,
  {Johnston}, K.~J., {Mozurkewich}, D., \& {Schmitt}, H. 2006{\natexlab{a}},
  \apj, 636, 1087

\bibitem[{{Peterson} {et~al.}(2006{\natexlab{b}}){Peterson}, {Hummel}, {Pauls},
  {Armstrong}, {Benson}, {Gilbreath}, {Hindsley}, {Hutter}, {Johnston},
  {Mozurkewich}, \& {Schmitt}}]{2006Natur.440..896P}
{Peterson}, D.~M., {Hummel}, C.~A., {Pauls}, T.~A., {Armstrong}, J.~T.,
  {Benson}, J.~A., {Gilbreath}, G.~C., {Hindsley}, R.~B., {Hutter}, D.~J.,
  {Johnston}, K.~J., {Mozurkewich}, D., \& {Schmitt}, H.~R. 2006{\natexlab{b}},
  \nat, 440, 896

\bibitem[{{Pont} {et~al.}(2009){Pont}, {Hebrard}, {Irwin}, {Bouchy}, {Moutou},
  {Ehrenreich}, {Guillot}, \& {Aigrain}}]{2009arXiv0906.5605P}
{Pont}, F., {Hebrard}, G., {Irwin}, J.~M., {Bouchy}, F., {Moutou}, C.,
  {Ehrenreich}, D., {Guillot}, T., \& {Aigrain}, S. 2009, ArXiv e-prints

\bibitem[{{Press} {et~al.}(1992){Press}, {Teukolsky}, {Vetterling}, \&
  {Flannery}}]{1992nrca.book.....P}
{Press}, W.~H., {Teukolsky}, S.~A., {Vetterling}, W.~T., \& {Flannery}, B.~P.
  1992, {Numerical recipes in C. The art of scientific computing} (Cambridge:
  University Press)

\bibitem[{{Robinson} {et~al.}(2007){Robinson}, {Laughlin}, {Vogt}, {Fischer},
  {Butler}, {Marcy}, {Henry}, {Driscoll}, {Takeda}, \&
  {Johnson}}]{2007ApJ...670.1391R}
{Robinson}, S.~E., {Laughlin}, G., {Vogt}, S.~S., {Fischer}, D.~A., {Butler},
  R.~P., {Marcy}, G.~W., {Henry}, G.~W., {Driscoll}, P., {Takeda}, G., \&
  {Johnson}, J.~A. 2007, \apj, 670, 1391

\bibitem[{{Russell}(1939)}]{1939ApJ....90..641R}
{Russell}, H.~N. 1939, \apj, 90, 641

\bibitem[{{Sato} {et~al.}(2008{\natexlab{a}}){Sato}, {Izumiura}, {Toyota},
  {Kambe}, {Ikoma}, {Omiya}, {Masuda}, {Takeda}, {Murata}, {Itoh}, {Ando},
  {Yoshida}, {Kokubo}, \& {Ida}}]{2008PASJ...60..539S}
{Sato}, B., {Izumiura}, H., {Toyota}, E., {Kambe}, E., {Ikoma}, M., {Omiya},
  M., {Masuda}, S., {Takeda}, Y., {Murata}, D., {Itoh}, Y., {Ando}, H.,
  {Yoshida}, M., {Kokubo}, E., \& {Ida}, S. 2008{\natexlab{a}}, \pasj, 60, 539

\bibitem[{{Sato} {et~al.}(2008{\natexlab{b}}){Sato}, {Toyota}, {Omiya},
  {Izumiura}, {Kambe}, {Masuda}, {Takeda}, {Itoh}, {Ando}, {Yoshida}, {Kokubo},
  \& {Ida}}]{2008PASJ...60.1317S}
{Sato}, B., {Toyota}, E., {Omiya}, M., {Izumiura}, H., {Kambe}, E., {Masuda},
  S., {Takeda}, Y., {Itoh}, Y., {Ando}, H., {Yoshida}, M., {Kokubo}, E., \&
  {Ida}, S. 2008{\natexlab{b}}, \pasj, 60, 1317

\bibitem[{{Seager} \& {Hui}(2002)}]{Seager.oblateness}
{Seager}, S. \& {Hui}, L. 2002, \apj, 574, 1004

\bibitem[{{Snellen} {et~al.}(2008){Snellen}, {Koppenhoefer}, {van der Burg},
  {Dreizler}, {Greiner}, {de Hoon}, {Husser}, {Kruhler}, {Saglia}, \&
  {Vuijsje}}]{2008arXiv0812.0599S}
{Snellen}, I.~A.~G., {Koppenhoefer}, J., {van der Burg}, R.~F.~J., {Dreizler},
  S., {Greiner}, J., {de Hoon}, M.~D.~J., {Husser}, T.~O., {Kruhler}, T.,
  {Saglia}, R.~P., \& {Vuijsje}, F.~N. 2008, ArXiv e-prints

\bibitem[{{Triaud} {et~al.}(2009){Triaud}, {Queloz}, {Bouchy}, {Moutou},
  {Collier Cameron}, {Claret}, {Barge}, {Benz}, {Deleuil}, {Guillot},
  {H{\'e}brard}, {Lecavelier des {\'E}tangs}, {Lovis}, {Mayor}, {Pepe}, \&
  {Udry}}]{2009arXiv0907.2956T}
{Triaud}, A.~H.~M.~J., {Queloz}, D., {Bouchy}, F., {Moutou}, C., {Collier
  Cameron}, A., {Claret}, A., {Barge}, P., {Benz}, W., {Deleuil}, M.,
  {Guillot}, T., {H{\'e}brard}, G., {Lecavelier des {\'E}tangs}, A., {Lovis},
  C., {Mayor}, M., {Pepe}, F., \& {Udry}, S. 2009, ArXiv e-prints

\bibitem[{{von Zeipel}(1924)}]{1924MNRAS..84..665V}
{von Zeipel}, H. 1924, \mnras, 84, 665

\bibitem[{{Winn} {et~al.}(2006){Winn}, {Johnson}, {Marcy}, {Butler}, {Vogt},
  {Henry}, {Roussanova}, {Holman}, {Enya}, {Narita}, {Suto}, \&
  {Turner}}]{2006ApJ...653L..69W}
{Winn}, J.~N., {Johnson}, J.~A., {Marcy}, G.~W., {Butler}, R.~P., {Vogt},
  S.~S., {Henry}, G.~W., {Roussanova}, A., {Holman}, M.~J., {Enya}, K.,
  {Narita}, N., {Suto}, Y., \& {Turner}, E.~L. 2006, \apjl, 653, L69

\bibitem[{{Winn} {et~al.}(2007){Winn}, {Johnson}, {Peek}, {Marcy}, {Bakos},
  {Enya}, {Narita}, {Suto}, {Turner}, \& {Vogt}}]{2007ApJ...665L.167W}
{Winn}, J.~N., {Johnson}, J.~A., {Peek}, K.~M.~G., {Marcy}, G.~W., {Bakos},
  G.~{\'A}., {Enya}, K., {Narita}, N., {Suto}, Y., {Turner}, E.~L., \& {Vogt},
  S.~S. 2007, \apjl, 665, L167

\bibitem[{{Wolf} {et~al.}(2007){Wolf}, {Laughlin}, {Henry}, {Fischer}, {Marcy},
  {Butler}, \& {Vogt}}]{2007ApJ...667..549W}
{Wolf}, A.~S., {Laughlin}, G., {Henry}, G.~W., {Fischer}, D.~A., {Marcy}, G.,
  {Butler}, P., \& {Vogt}, S. 2007, \apj, 667, 549

\end{thebibliography}

\newpage

\end{document}